\documentclass[a4paper,twoside]{mn2e}
\usepackage{rotate}
\usepackage{epsf}
\newcommand{\R}{\cal R}
\newcommand{\Rns}{{\cal R}_{ NS}}

\newcommand{\M}{\cal M}

\newcommand{\myfrac}[2]{\left(\frac{#1}{#2}\right)}   
\newcommand{\NUlim}{\nu_{lim}}

\title{Gravitational wave background from coalescing compact stars
in eccentric orbits}
\author
[V.B. Ignatiev et al.]
{V.B. Ignatiev$^1$, A.G.Kuranov$^{2}$, K.A. Postnov$^{1,3}$,
M.E. Prokhorov$^2$
\\
\\
\\
$^1$ Faculty of Physics, Moscow State University, 119899 Moscow, Russia\\
$^2$ Sternberg Astronomical Institute, 
119899 Moscow, Russia\\
$^3$  Max-Planck Institut f\"ur
Astrophysik, 85740 Garching, Germany}

\date{Accepted 2001...
      Received 2001 ...}

\begin{document}
\maketitle
\label{firstpage}

\begin{abstract}

Stochastic gravitational wave background produced by a stationary 
coalescing population of binary neutron stars in the Galaxy is calculated.
This background is found to constitute a confusion limit within the LISA 
frequency band up to a limiting frequency $\NUlim{}\sim 10^{-3}$~Hz, leaving
the frequency window $\sim 10^{-3}$--$10^{-2}$~Hz open for the potential 
detection of cosmological stochastic gravitational waves and of signals 
involving massive black holes out to cosmological distances.

\end{abstract}
%
\section{Introduction}

In a short time, gravitational wave astronomy is expected to start
collecting first data from different astrophysical sources of gravitational
radiation, most probably from coalescing binary neutron stars and black
holes (for a recent review, see Grishchuk et al. 2001). The first large
gravitational wave (GW) interferometers LIGO, VIRGO, GEO-600 and TAMA-300
will be sensitive to dimensionless metric variations in the frequency range
10-1000 Hz at a level of $h\sim 10^{-21}$ (Braginsky 2000 and references
therein). Due to seismic noise limitations lower frequencies 
($10^{-4}-10^{-1}$ Hz) can be studied only with
space-born antennas, such as LISA laser interferometer (Bender et al.
2000).
Prospects for finding astrophysical sources within the LISA frequency band are
very good, especially for supermassive binary black holes which can be
observed by LISA with a record signal-to noise ratio of order 1000 from
cosmological distances (Vecchio et al. 1997).

In addition to signals from individual sources, GW interferometers can
detect stochastic gravitational waves. Such waves (GW backgrounds, or GW
noises) can be produced by a large collection of unresolved individual
sources, e.g. by binary stars (Mironovskij 1965, Lipunov and Postnov 1987,
Bender and Hils 1997, Kosenko and Postnov 1998) 
or rapidly rotating neutron stars (Giazotto et al.
1997, Postnov and Prokhorov 1997, Ferrari et al. 1999). These GW backgrounds
are usually considered as unwanted additional noises to the intrinsic noises
of GW detectors, since they could potentially mimic stochastic GW
backgrounds of cosmological origin (primordial or relic GW) that bring a
valuable information about physical processes in the very early Universe
(Grishchuk et al. 2001 and references therein).

Ordinary galactic binaries, in which at least one of the component
is a normal main-sequence star,   
mostly contribute at low-frequency band ($\nu \sim 10^{-7}-10^{-5}$ Hz)
(Mironovskij 1965).
At higher frequencies $10^{-4}-10^{-2}$ Hz compact coalescing 
white dwarf (WD) binaries should form a noticeable
background above the LISA noise curve (Lipunov and Postnov 1987, 
Bender and Hils 1997, Grishchuk et al. 2001). Generally, 
the level of the GW noise within a fixed frequency interval 
from a collection of $N$ unresolved independent  
sources in terms of dimensionless amplitude is $h\propto \sqrt{N}$.
In case of binaries that loose the orbital angular momentum 
due to gravitational radiation back reaction, this level is 
$h\propto \sqrt{\R}$, where $\R$ is the coalescing rate of 
these binaries (see Grishchuk et al. 2001 for more detail). 

Another important quantity is 
the upper frequency $\NUlim{}$ above which  
each individual source can be resolved during one-year observation time
(i.e. within the frequency bin $\Delta \nu=3\times 10^{-8}$ Hz). 
In fact, it is this limiting frequency which mostly relate to 
the challenging task of detection of a relic GW background: 
at $\nu>\NUlim{}$ we will be able in principle to single out
individual sources and thus have prospects to register 
cosmological GW noise using one interferometer (Grishchuk
et al. 2001).                                    

The obvious condition for $\NUlim{}$ reads
${\R} \Delta T(\NUlim{}) = 1$, where $\Delta T(\nu)$ is the time each 
source `spends` within the frequency bin at a given frequency $\nu$.
For example, in the simplest case of a collection of galactic binary WD
in circular orbits which coalesce at a constant rate 
${\R}={\R}_{300}\times(1/300)$~yr$^{-1}$

\begin{equation}
\begin{array}{rl}
\label{intro}
\NUlim{}(\hbox{WD})&\approx (1.2\times 10^{-3}\hbox{Hz})~{\R}_{300}^{3/11}\nonumber\\ 
&\times
\myfrac{\Delta \nu}{3\times 10^{-8}\hbox {Hz}}^{3/11}\myfrac{\M}
{0.52 M_\odot}^{-5/11}
\end{array}
\end{equation}
where $\M$ is the chirp mass of the binary.
For merging binary neutron stars (NS) $\NUlim{}$ 
calculated as above would give even smaller
value, $\NUlim{}($NS$)=3\times 10^{-4}$ Hz, since the binary NS coalescence rate
in most optimistic scenarios is $\Rns{}\sim 10^{-4}$ yr$^{-1}$
(Grishchuk et al. 2001, Kalogera et al. 2000).
Note that the uncertainty in this rate by a factor of 3 or so 
is of minor importance since $\NUlim{}\propto {\R}^{3/11}$.

However, there is an important difference between the merging WD
and NS. Binary WD must have almost circular orbits from the very beginning
since they result from a spiral-in process during the common envelope stage.
Unlike binary WD, binary NS must have (and this is what we actually observe 
in the known binary radiopulsars with NS companions) 
a significant eccentricity at birth, 
since they are formed after two supernova explosions with substantial
mass loss from the system. The possible 
asymmetry of supernova explosion leading to the natal kick velocity 
of newborn NS additionally affects the orbital parameters leading
to higher orbital eccentricities.      

It is well known (Peters and Mathews 1963)
that
an eccentric binary system emit GW in a wide frequency band. This means that
the high-order harmonics from an eccentric binary system should be observed
at frequencies $\nu > 2\nu_K$ ($2\nu_K$ is twice the Keplerian orbital
frequency at which a circular binary radiates GW), so the entire population
of galactic eccentric binary NS should contribute in a broader frequency band
than analogical circular binaries would do. The effect may be not small,
since the total number of NS+NS binaries in the Galaxy is quite large, as
the simple estimate below shows. Assume the stationary formation rate
${\R}_f$ of NS binaries in the Galaxy (which is a reasonable approximation
during at least the last 5 billion years). Then the total number of them is
of order ${\R}_{f}\times T_{gal}$, where $T_{gal}$ is the galactic age. Note that
the actual NS formation rate is higher than the NS merging rate $\Rns{}$
since a NS
binary could merge only if its initial semi-major axis $a_i$ and
eccentricity $e_i$ satisfy the condition
$T_{coal}(a_i,e_i)<t_{gal}$, where $T_{coal}$ is the time the systems takes 
from the formation to coalescence due to GW orbit decay.
So the minimum number of NS binaries in the Galaxy is estimated 
to be of order $\Rns{}\times T_{gal}= 10^{-4}\times 10^{10}\sim 10^6$.

To calculate the GW background produced by binary NS, the following steps
should be done. (1) NS binaries form with some initial distribution over
orbital semi-major axes and eccentricities $F_{in}(a,e)$. This distribution can
be found from evolutionary calculations using e.g. binary population
synthesis method (Lipunov et al. 1996). (2) Next, if orbital parameters of
such binaries evolve only due to GW emission, a stationary distribution
function $F_{st}(a,e)$ can be recovered (Buitrago et al. 1994). (3) Knowing
the GW spectrum of a binary with given $(a,e)$  
(Peters and Mathews 1963), the final GW
noise from these sources can be computed.

In the present paper we specially focus on the effect of eccentricity of
coalescing NS binaries on the level and limiting 
frequency of the stochastic GW background they form. We find, as expeted,
that the level of this background is lower than from coalescing
galactic binary WD, and the limiting frequency $\nu_{lim}(e\ne 0)$ is
higher than calculated for circular case.  
The initial (model-dependent) distribution of NS binaries
only slightly affects the result. This leaves the frequency interval 
$\sim 10^{-3}-10^{-2}$~Hz open for searches of primordial
GW backgrounds.
\vspace{-4mm}

\section{Initial galactic binary NS distribution}

\begin{figure}
\centerline{\epsfysize=1\hsize
\epsfbox{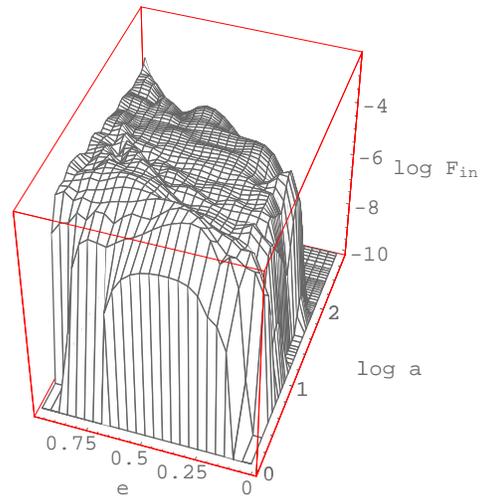}}
\caption{The initial binary NS distribution $F_{in}(a,e)$ for the kick velocity 
amplitude $v_{k}=200$ km/s. Initial semimajor axis $a$ is in solar units.
}
\label{fin}
\end{figure}

To calculate the initial binary NS distribution $F_{in}(a,e)$ in the Galaxy
we used the population synthesis method described by many authors
(Lipunov, Postnov, Prokhorov 1996; Lipunov, Postnov, Prokhorov 1997; 
Portegies Zwart, Yungelson 1998).
It uses the simulation of evolution of a large number of
binaries with initial parameters (masses of the components,
semimajor axes, eccentricities etc.) distributed according to 
some (taken from observations or model) laws. There are also 
evolutionary parameters, such as efficiency of common envelope stage,
kick velocity during supernova explosion, initial spin of compact objects,
etc. (see Grishchuk 2001 for more detail), of which kick velocity 
imparted to compact object (neutron star or black hole) at birth
mostly impacts the resulting distribution $F_{in}(a,e)$. 
We assumed a Maxwellian form of the kick velocity distribution and
varied its amplitude $v_k$ from 0 to 400 km/s.
The resulting normalized initial distribution $F_{in}(a,e)$ for 
$v_k=200$~km/s is shown 
in Fig.~\ref{fin}.

NS binaries are formed in a very broad interval of $a$ and $e$, but
interesting for us here will be only those that can coalesce over 
the galactic age $T_{coal}<T_{gal}$, 
since only such binaries can form a stationary distribution.
It is seen from Fig.~\ref{fin} that most binaries form with small or
moderate eccentricities ($e<0.5$) 
in a wide range of semimajor axes $a_{in}\sim 1-300
R_\odot$. Increase in the kick velocity amplitude broadens the
initial distribution function over eccentricity.
\vspace{-5mm}

\section{Stationary binary NS distribution}

GW back reaction causes orbital decay of binary system. For an eccentric binary
star, this
can be described in quadrapole approximation by ordinary differential 
equations for orbital semimajor axis and eccentricity (Peters~\& Mathews, 1963):

\begin{equation}
\label{da/dt}
\frac{da}{dt} =- \frac{64}{5}
\frac{G^3m_1m_2(m_1+m_2)}{c^5a^3(1-e^2)^{7/2}}(1+\frac{73}{24}e^2+\frac{37}{96}e^4)
\end{equation}

\begin{equation}
\label{de/dt}
\frac{de}{dt}=
\frac{304}{15}\frac{G^3m_1m_2(m_1+m_2)e}{c^5a^4(1-e^2)^{5/2}}(1+\frac{121}{304}e^2)
\end{equation}

Given the initial function $F_{in}(a,e)$, it is
straightforward to 
find a stationary distribution function $F_{st}(a,e)$
(Buitrago et al. 1994):

\begin{equation}  
\begin{array}{r@{}l} 
\label{fst}
F_{st}(a,e)=\,&
\frac{\displaystyle 15c^5{\R}_f}{\displaystyle 304G^3m_1m_2(m_1+m_2)}
\\[2mm]
&\times\frac{\displaystyle a^4(1-e^2)^{7/2}}{\displaystyle e^{31/19}
(1+\frac{\displaystyle 121}{\displaystyle 304}e^2)^{3169/2299}}
\\[3mm]    
&\times\int\limits_{0}^{1}\frac{\displaystyle z^{12/19}
(1+\frac{\displaystyle 121}{\displaystyle 304}z^2)^{870/2299}}
{\displaystyle 1-z^2}F_{in}(a(z),z)\,dz 
\end{array} 
\end{equation}

\begin{figure} 
\centerline{\epsfysize=0.8\hsize
\epsfbox{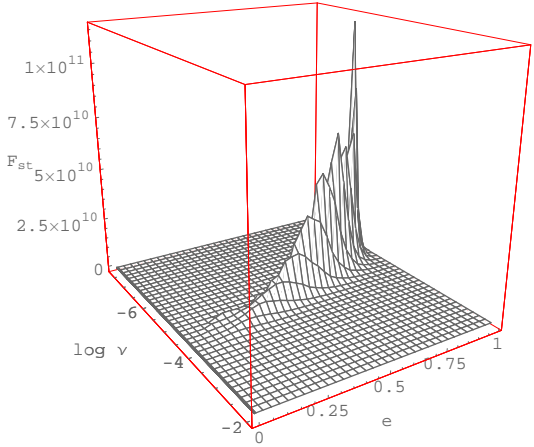}}
\caption{The stationary distribution function $F_{st}$ for binary
NS in eccentric
orbits which coalesce due to GW emission in a 
time interval shorter than the galactic age $10^{10}$ years.}
\label{ttt6}
\end{figure}

At present, only some fraction of systems from the initial
distribution with $T_{coal}<T_{gal}$ reaches stationarity. The function
$F_{st}(\nu_K,e)$ is calculated for this part of $F_{in}(a,e)$
from Fig.~\ref{fin} and is shown in Fig.~\ref{ttt6}. In this figure we use
the Keplerian frequency $\nu_{K}=\sqrt{(1/2\pi)G(m_1+m_2)/a^3}$
instead of the semi-major axis $a$.

\section{GW spectrum from non-circular binary stars}

In the simplest case of two point masses in a circular orbit, the energy
losses caused by the quadrupole gravitational wave emission 
$L_0=\frac{\displaystyle dE}{\displaystyle dt}(\nu_{K})$ are
defined by (Einstein, 1916, 1918; Landau \& Lifshiz 1971):
\begin{equation}
\label{lo}
L_0= \frac{32}{5} \frac{G^4}{c^5}\frac{m_1^2m_2^2(m_1+m_2)}{a^5}
\end{equation} 
where $G$ is Newton's gravitational constant, $c$ is the speed of the light in a vacuum and
$m_1, m_2$ are the masses of stars.
All GW radiation in this case is emitted at a single frequency
$2\nu_{K}$.

For a non-circular orbit with eccentricity $e$ the GW luminosity
increases (Peters \& Mathews 1963):
\begin{equation}
\label{le}
\frac{dE}{dt}=
L_0\times\frac{1+\frac{\displaystyle 73}{\displaystyle 24}e^2
+\frac{\displaystyle 37}{\displaystyle 96}e^4}{(1-e^2)^{7/2}}
\end{equation} 
The emission now is widely spread over higher-order harmonics to the main
frequency  (Peters \& Mathews 1963):
\begin{equation}
\label{le2}
\frac{dE}{dt}(\nu=n\nu_{k})=L_0
\times g(n,e)
\end{equation}
\begin{eqnarray}
\label{g(n,e)}
g(n,e)= \frac{n^4}{32}([J_{n-2}(ne)-2eJ_{n-1}(ne)+\frac{2}{n}J_{n}(ne)\nonumber\\
+2eJ_{n+1}(ne)-J_{n+2}(ne)]^2+(1-e^2)[J_{n-2}(ne)\nonumber\\
-2J_{n}(ne)+J_{n+2}(ne)]^2+\frac{4}{3n^2}{J_{n}}^2(ne))
\end{eqnarray}
Here $J_n$ is Bessel function and $n$ is the number of the harmonics.

The total energy emitted per second 
in a frequency interval $\Delta\nu$
by all binaries in the Galaxy is
\begin{equation}
\begin{array}{l}
\label{de/dt/dnu}
\frac{\displaystyle dE}{\displaystyle dt} 
=\int\limits_{\nu}^{\nu+\Delta\nu}\biggl(
~\sum\limits_{n=1}^\infty \int\limits_0^1
L_0(\frac{\displaystyle \nu}{\displaystyle n})
\, g(n,e)
\,F_{st}(\frac{\displaystyle \nu}{\displaystyle n},e)\, de
\biggr)\,\frac{\displaystyle d\nu}{\displaystyle n}
\\
\approx\biggl(~\sum\limits_{n=1}^{n_{lim}} \int\limits_0^1
L_0(\frac{\displaystyle \nu}{\displaystyle n})
\,g(n,e)
\,F_{st}(\frac{\displaystyle \nu}{\displaystyle n},e)\, de\biggr)
\frac{\displaystyle \Delta\nu}{\displaystyle n}
\\
\end{array}
\end{equation}
where $F_{st}(\frac{\displaystyle \nu}{\displaystyle n},e)
\equiv F_{st}(\nu_{k},e)$ is a stationary
distribution function of binary NS.
In the last equation we take into account $\Delta \nu \ll \nu$.
The amplitude of high-order harmonics rapidly decreases, so we stopped 
the summation of harmonics for $n>n_{lim}$, 
$n_{lim}: g(n_{lim},e)= \epsilon \times \max_n[g(n,e)]$.
We assumed $\epsilon=10^{-4}, 10^{-6}, 10^{-10}$. Decreasing 
$\epsilon$ increases the total number of harmonics which 
contribute to the given frequency bin, but practically 
does not change the number of the most powerful harmonics 
within it (see Fig. 7). 
\vspace{-3mm}

\section{Stationary stochastic GW noise from coalescing binary NS}
\begin{figure}
\centerline{\epsfysize=1\hsize
\rotate[r]{
\epsfbox{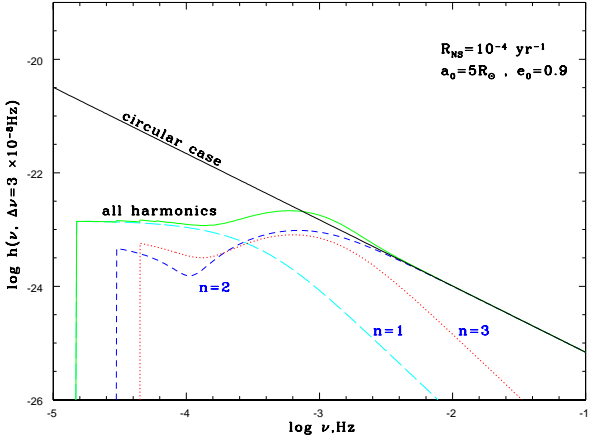}}}
\caption{The resulting GW 
spectrum and the contribution from the first, second and third  harmonics.
All systems form with~$a_0=5 R_{\odot},e_0=0.9$. 
For comparison we show the spectrum for circual NS binaries
coalescing with the same rate, and with initial orbital periods longer 
than $2\times 10^5$s.}
\label{delta1}
\end{figure}

\begin{figure}
\centerline{\epsfysize=1\hsize
\rotate[r]{
\epsfbox{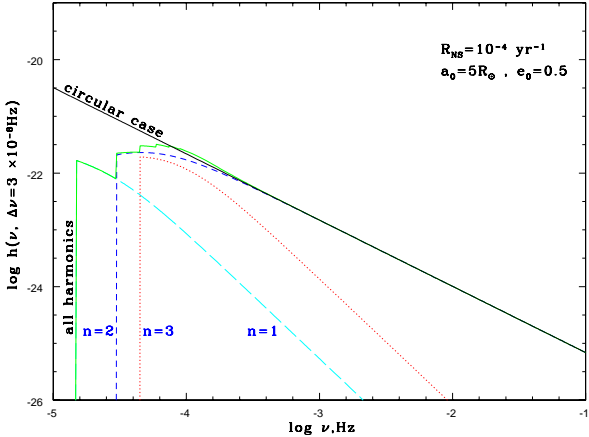}}}                                      	
\caption{The same as in Fig~\ref{delta1} for 
systems with $a_0=5 R_{\odot},e_0=0.5$.}
\label{delta2}
\end{figure}

First, consider the GW spectrum from a delta-function initial distribution:
$F_{in}(a,e)=\delta(a-a_0,e-e_0)$, that is assume all binaries to form at one
point $(a_0,e_0)$ in $(a-e)$ configuration space at a rate of
$\Rns{}=10^{-4}$ yr$^{-1}$. The spectrum formed by the first three
harmonics and the total stationary GW background for this case are shown in
Fig.~\ref{delta1} and Fig.~\ref{delta2} for $a_0=5 R_\odot$ and
$e_0=0.5,~0.9$, respectively. It is seen that increase in the initial 
eccentricity strongly affects the shape of the spectrum up to some frequency
at which the eccentricities of binaries in the stationary distribution 
become sufficiently small. 
Above this frequency only the second harmonics from almost circular 
binaries contributes to the total spectrum. 
The non-monotonic dependence of the spectrum formed by the harmonics is
due to the non-monotonic behaviour of the energy emitted at each harmonics 
with eccentricity (Eq.~(\ref{le2})-(\ref{g(n,e)})).

Now we turn to the expected GW stochastic background from  
galactic NS binaries. We shall
assume all the sources to lie at one distance $r=7.9$~kpc, which is close to
the average distance to stars in our Galaxy. This simplifying assumption
is unlikely to change our general conclusions.
At each frequency $\nu$ we sum up the GW flux from all the harmonics that fall 
within the chosen frequency bin $\Delta\nu=3\times 10^{-8}$~Hz from 
lower-frequency non-circular systems in the calculated stationary distribution
$F_{st}(\nu,e)$ (Eq.~(\ref{fst})), as modified
by including only the part that has come to equilibrium in the    
lifetime of the galaxy, as discussed earlier.

\begin{figure}
\centerline{\epsfysize=1\hsize
\rotate[r]{
\epsfbox{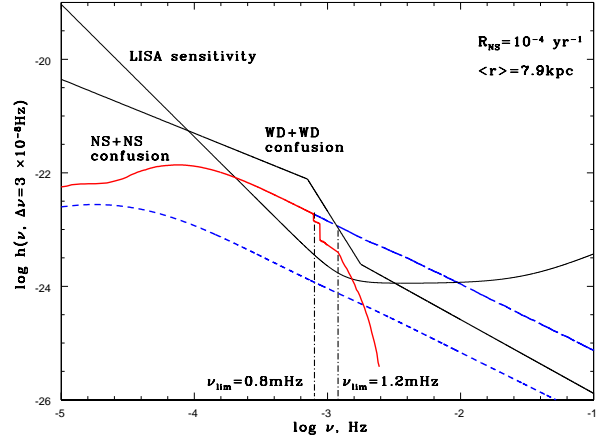}}}
\caption{GW background from stationary binary NS population  
calculated for a model spiral 
galaxy with the NS+NS coalescing rate ${\cal R}_{NS}=10^{-4}$~yr$^{-1}$.
The limiting frequecy above which 
the most powerful harmonics can be resolved in one-year observation
is  $\NUlim{}\simeq 0.8$~mHz. If no harmonics were subtracted from the
frequency bin $3\times 10^{-8}$~Hz, 
the background would continue toward higher frequencies 
as shown by the long-dashed line. THe lower dashed line shows
contribution from extragalactic NS+NS binaries.
The upper solid line schematically represents WD+WD confusion limit 
from Bender and Hils (1997) ($\NUlim{}\simeq 1.2$~mHz, Eq. (1)). 
The proposed LISA sensitivity 
is from Thorne (1995) [Fig. 14]}
\label{spectr}
\end{figure}

The resulting noise curve is shown in Fig. \ref{spectr} 
in terms of dimensionless
amplitude $h$ 
\begin{equation}
h^2(\nu)= \frac{G}{c^3 r^2 (\pi \nu)^2} \frac{dE}{dt}
\end{equation} 
As expected, the NS+NS confusion noise lies below WD+WD curve, mainly due to 
lower $\Rns{}$. High-order harmonics from non-circular NS binaries mostly 
contribute at lower frequencies, 
so starting from $\nu\sim 10^{-4}$~Hz the calculated 
noise curve practically coincides with that formed by circular NS binaries 
coalescing with the same rate $\Rns{}=10^{-4}$~yr$^{-1}$. 

\begin{figure}
\centerline{\epsfysize=1\hsize
\rotate[r]{
\epsfbox{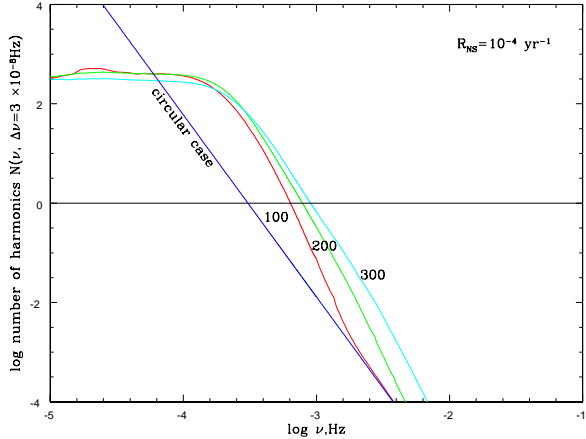}}} 
\caption{Number of harmonics from non-circular galactic NS binaries
producing the stochastic noise 
in the frequency bin $\Delta \nu=3\times10^{-8}$~Hz as a function of
the kick velocity. The curves are labeled with
kick velocity amplitudes $100, 200$ and $300$ km/s assumed in 
the population synthesis calculations.
The straight line is for the circular NS binaries.
The limiting frequency $\nu_{lim}$ above which individual
harmonics can be resolved follows from the condition
$N(\nu_{lim},\Delta\nu)=1$.}
\label{num}
\end{figure}

\begin{figure}
\centerline{\epsfysize=1\hsize
\rotate[r]{
\epsfbox{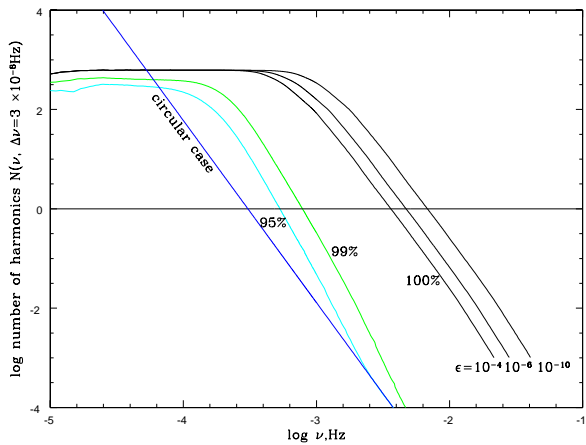}}} 
\caption{Number of the most powerful 
harmonics from non-circular galactic NS binaries which contribute 
95\%, 99\% and 100\% of the total GW energy 
collected in the bin $\Delta\nu=3\times 10^{-8}$
Hz. The effect of 
increase in the number of harmonics $n_{lim}$ determining
the total energy collected in the bin (Eq. (9)) is shown 
for the total number of harmonics (100\%) for different $\epsilon$. 
Kick velocity amplitude $200$ km/s.}
\label{num1}
\end{figure}

More important in the non-circular case is the increase of the limiting 
frequency $\NUlim{}$ above which individual 
sources can be resolved during a one-year observation.
To estimate it, we calculate the number of
harmonics inside the frequency bin $\Delta\nu$.
Formally, this number is more than one at any frequency, 
but at high frequencies the main contribution must come
from circular binaries (2-nd harmonics), the total GW power 
from higher-order harmonics from low-frequency eccentric 
systems becoming gradually less and less.  
So the number of harmonics 
contributing in a given frequency bin $\Delta\nu=3\times 10^{-8}$~Hz
were counted, starting with 
the strongest one and continuing until 99
bin~(Eq.(\ref{de/dt/dnu})) had been included.
The number of such harmonics as a function of frequency and the 
assumed kick velocity amplitude 
$N(\nu,\Delta\nu)$ is shown in Fig.~\ref{num}. The limiting frequency 
$\NUlim{}$ is determined from equation $N(\NUlim{},\Delta\nu)=1$.
To within uncertainties of our calculations ($\Rns{},F_{in}(a,e)$,
etc.), 
$\NUlim{}\simeq 10^{-3}$~Hz, close to the break in the 
confusion limit from binary WD. 
For comparison, in the same Fig.~\ref{num} we 
show $N(\nu,\Delta\nu)$ for stationary circular NS binaries coalescing with
the same rate $\Rns{}=10^{-4}$~yr$^{-1}$ 
($\NUlim{}({e=0})\propto 3\times 10^{-4}$~Hz).
The effect of increasing  the chosen level for the GW power
from the strongest harmonics in the bin is 
illustrated in Fig. \ref{num1}. Increasing the level from 95\%
to 100\% 
changes the  limiting frequency  $\NUlim{}$ 
by almost an order of magnitude.  

This procedure, however, is not complete. Indeed, 
consider a frequency bin next to thus defined $\nu_{lim}$. 
We may ask the question which harmonics will be most
probably absent inside it. The answer is those which 
are the least probable at this frequency. 
The probability to find a harmonics in the bin
at a given frequency is determined by the number 
of the harmonics and the stationary distribution function 
of sources $f(\nu,e)$ 
It happens that this is harmonics number one (mostly due to
a steep decrease of the systems' stationary distribution function). 
If we remove all 1st harmonics of systems with orbital frequencies
falling into the chosen bin from  
the sum~Eq.(\ref{de/dt/dnu}), we are left again with some 
(smaller) GW power in the bin and may wish to find the limiting
frequency exactly in the same way as described above
(i.e. by fixing some level and summing up the strongest harmonics
up to this level), call it $\nu_{lim,-1}$. Above this new 
limiting frequency we can repeat the entire procedure
to find $\nu_{lim, -2}$ (this time the 2d harmonics is 
least probable to be found in the bin next to the new limiting
frequency), etc. until the noise level of the
detector is reached (it is less interesting for us to 
see the changes in the spectrum below it). 
The step-like line which continues 
the spectrum above $\sim 1$ mHz 
in Fig. \ref{spectr} illustrates this
procedure. This line of course do not represent the "real"
GW noise curve from binary NS and just give a feeling 
of how it most probably behaves at $\nu_{lim}$,
above which individual harmonics from coalescing binary NS 
can be singled out.  

Extragalactic NS binaries would also form an isotropic confusion noise, but
the level of extragalactic GW backgrounds is generally an order of magnitude
lower than the galactic one (e.g. calculations of Kosenko and Postnov, 1998)
and is beyond reach by the expected LISA sensitivity (the bottom dashed
curve in Fig. 5). The limiting frequency $\NUlim{}$ for extragalactic NS
binaries finds from the condition for circular systems and can be as higher
as $\sim 0.2$ Hz (Ungarelli and Vecchio, 2000).
\vspace{-4mm}

\section{Conclusions}

We studied the effect of non-circularity of realistic coalescing NS 
binaries on stochastic galactic background formed by  stationary NS+NS
distribution $F_{st}(\nu,e)$. The level of the background is 
found to be by one order of magnitude less than from 
coalescing binary WD systems, in correspondence with
the lower NS binary coalescence rate.
The limiting frequency above which individual harmonics 
can be resolved at a 99\% level over this noise
is $\NUlim{}\approx 0.8 \times 10^{-3}$~Hz, close to $\NUlim{}$ for coalescing
binary WD. 
Detection of an isotropic stochastic GW signal by LISA at higher frequency 
would strongly indicate its cosmological origin.
So our study confirms that within the frequency range $10^{-3}$--$10^{-2}$~Hz
there are prospects to detect cosmological stochastic GW by means of
one space interferometer LISA as suggested in Grishchuk et al 2001.
\vspace{-4mm}

\section*{Acknowledgments}
The authors thank L.P. Grishchuk for useful discussion and partial support 
from RFBR grants 00-02-17164 and 99-02-17884-a. 
AGK acknowledges RFBR for support through grant 00-02-17164. 
KAP acknowledges MPA f\"ur Astrophysik
(Garching) for hospitality.


\end{document}